\documentclass[preprint,showpacs,preprintnumbers,amsmath,amssymb]{revtex4}
\usepackage{dcolumn}
\usepackage{graphicx}
\usepackage{graphicx,amssymb,amsmath}
\usepackage{graphics}
\usepackage{ulem}
\usepackage{color}
\begin{document}

\title{What drives amyloid molecules to assemble into oligomers and fibrils?}
\author{Jeremy D.\ Schmit}
\email[]{schmit@maxwell.ucsf.edu}

\author{Kingshuk Ghosh}

\author{Ken Dill}
\email[]{dill@maxwell.ucsf.edu}
 \affiliation{Department of Pharmaceutical Chemistry, University of California,
 San Francisco, California 94158}

\begin{abstract}
We develop a general theory for three states of equilibrium of
amyloid peptides: the monomer, oligomer, and fibril.  We assume that
the oligomeric state is a disordered micelle-like collection of a
few peptide chains held together loosely by hydrophobic interactions
into a spherical hydrophobic core.  We assume that fibrillar amyloid
chains are aligned and further stabilized by `steric zipper'
interactions -- hydrogen bonding and steric packing, in addition to
specific hydrophobic sidechain contacts.  The model makes a broad
set of predictions, consistent with experiments:  (i)  Similar to
surfactant micellization, amyloid oligomerization should increase
with bulk peptide concentration.  (ii) The onset of fibrillization
limits the concentration of oligomers in the solution. (iii) The
average fibril length \emph{vs.} monomer concentration agrees with
data on $\alpha$-synuclein, (iv) Full fibril length distributions
follow those of $\alpha$-synuclein, (v) Denaturants should `melt
out' fibrils, and (vi) Added salt should stabilize fibrils by
reducing repulsions between amyloid peptide chains. Interestingly,
small changes in solvent conditions can: (a) tip the equilibrium
balance between oligomer and fibril, and (b) cause large changes in
rates, through effects on the transition-state barrier. This model
may provide useful insights into the physical processes underlying
amyloid diseases.
\end{abstract}

\maketitle

\section{Introduction}

What are the forces that stabilize aggregates of amyloid peptide
molecules?  This question is of interest because of the putative
role played by amyloid aggregation in diseases such as Alzheimer's,
Parkinson's, Mad Cow and type II diabetes \cite{Dobson:2003}.
Amyloid appears to aggregate into at least two different states:
\emph{amyloid oligomers}, which are small few-chain soluble
disordered clusters, and \emph{fibrils}, which are long many-chain
highly structured $\beta$-sheet-like aggregates.  The view has
recently emerged that the oligomers may be the toxic species, not
the fibrils, as had been expected because of the appearance of
plaques in disease \cite{Hardy:2002}. It has been challenging to
understand the physical principles of amyloid aggregation, in part
because of a lack of reductionist experimental model systems.  In
this breach, we believe that simple models can help guide and
interpret experiments.

The first challenge is to disentangle how much of amyloid formation
can be explained by equilibrium \emph{vs.} kinetics.  Often kinetics
is easier to study experimentally because measuring rates does not
require finding conditions of reversibility.  Yet, there is
experimental evidence of multiple stable states: monomers,
oligomers, and fibrils, and perhaps a precursor to the fibrillar
state, called the protofilament \cite{Chimon:2007}.  We believe
insights can be gained from first understanding the underlying phase
equilibria.

\section{Modeling the states of amyloid aggregation}

\begin{figure}[ht]
\vspace{0.6 cm}
\begin{center}
\includegraphics[angle=0,scale=1]{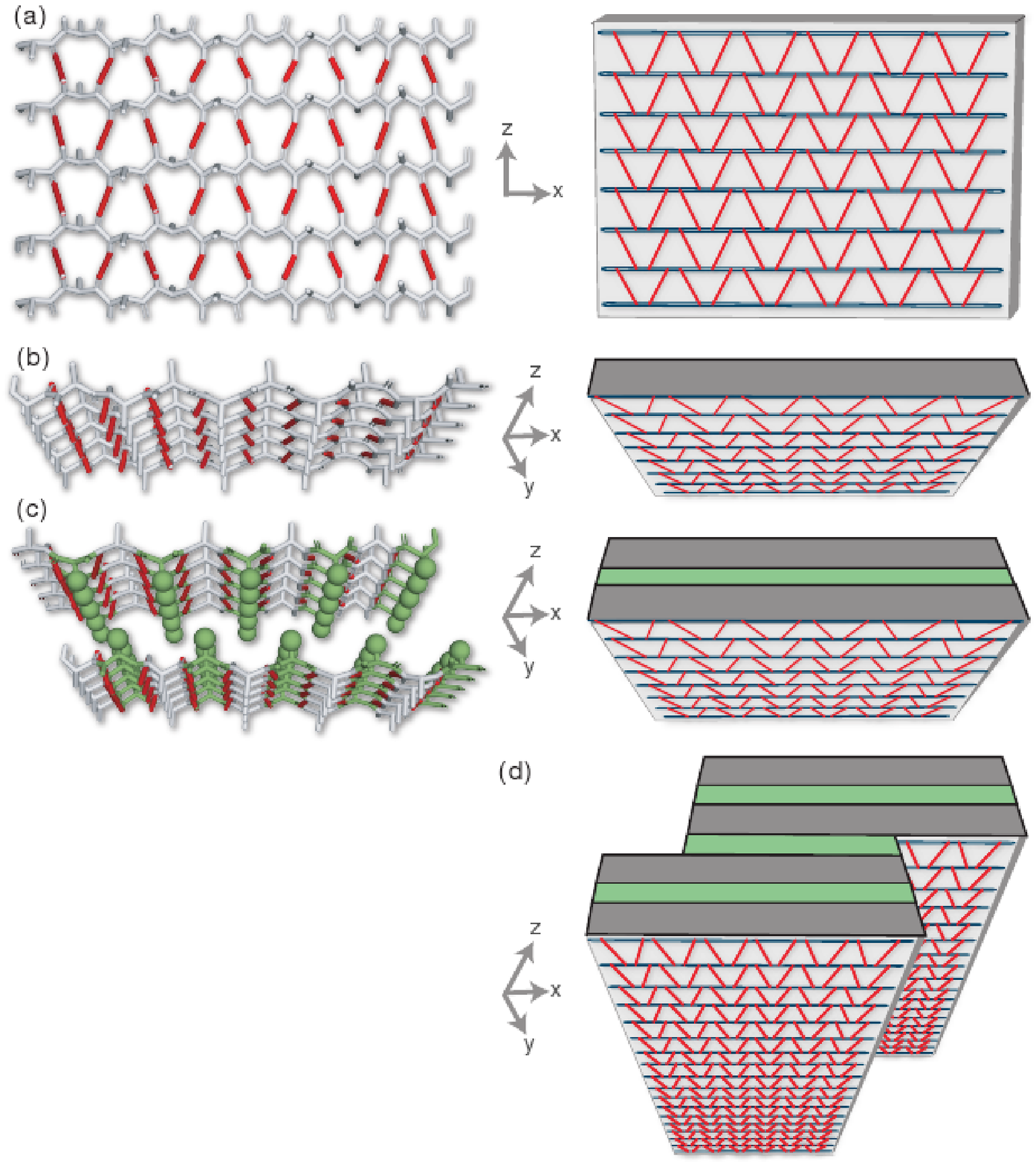}
\end{center}
\caption{ \label{fig:key} Assembly hierarchy of amyloid fibrils
shown in atomistic cartoon representation (left) and schematically
with $\beta$-sheets as blocks (right).  a) A single $\beta$ sheet
comprised of parallel $\beta$-strands. b) $\beta$-sheet observed
along fibrillization axis. c) Assembled $\beta$-sandwich
(protofilament) consisting of two $\beta$-sheets.  Note the steric
zipper interactions shown as interdigitating side chains (left) and
as a green layer (right). d) Mature fibril consisting of $p=2$
protofilaments.}
\end{figure}

We develop here a model of the equilibrium among the following
states: (a) isolated monomeric amyloid peptide molecules in
solution, (b) few-chain noncovalent aggregates \emph{(oligomers)} of
amyloid peptide molecules, (c) the single `macroscopic thread'
called a \emph{protofilament}, which is a noncovalent ordered
assembly of many chains, and (d) the \emph{fibril}, which is a
bundle of protofilaments.  These states are shown schematically in
Figs. \ref{fig:key} and \ref{fig:sandwich}.  (In this paper,
`monomer' refers to an individual peptide chain, not to a single
amino acid in a chain; see Fig. \ref{fig:sandwich}a)
\cite{Chen:2006,Chimon:2007,Glabe:2008}. Our interest here is in
peptides, such as A$\beta$, $\alpha$-synuclein, and IAPP, that do
not have single-chain native folded structures, so our model below
neglects any possible additional equilibria with a native folded
structure.

\subsection{Modeling the oligomer state}

\begin{figure}[ht]
\vspace{0.6 cm}
\begin{center}
\includegraphics[angle=0,scale=0.9]{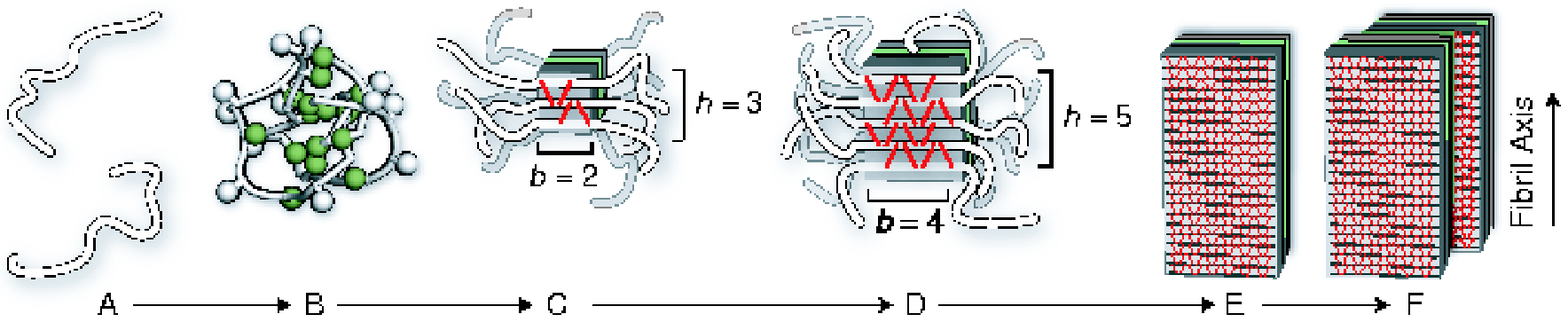}
\end{center}
\caption{ \label{fig:sandwich} Model of amyloid aggregation
equilibria.  Each black line indicates the peptide backbone.  Each
red line shows one hydrogen bond.  A) Isolated peptide monomers in
solution. B) Oligomeric assembly of a few peptide chains. C) Nucleus
of $\beta$-sheet structure.  The peptide backbone runs perpendicular
to the fiber axis.  D) Post-critical nucleus structure having more
$\beta$-structure. E) \emph{Protofilament} is a single long thread
of $\beta$-structure consisting of a \emph{$\beta$-sandwich}, two
$\beta$-sheet planes face-to-face.  F) \emph{Full fibril}, a bundle
of protofilaments, shown here containing $p=2$ protofilament
threads.}
\end{figure}

The oligomer state is shown in Fig. \ref{fig:sandwich}a.  For
A$\beta$ aggregates, oligomers range in size from trimers to
hundreds of monomers \cite{Chen:2006,Chimon:2007,Glabe:2008}.  We
model the amyloid oligomer state as a disordered spherical globule
having a hydrophobic core containing $N$ peptide chains. Each chain
has $L$ amino acids.  We approximate the free energy of
oligomerization, $\Delta F_{\mathrm{oligo}}$, from state $A$ to $B$
in Figure \ref{fig:sandwich}, in terms of the transfer of the $NL$
amino acids from water into the oligomeric core as
\cite{Dill:1985,Dill:1995},
\begin{equation}\label{eq:Foligo}
\Delta F_{\mathrm{oligo}} = \frac{\Delta F_{\mathrm{AB}}}{k T} =-\chi N L,
\end{equation}
where $kT\chi$ is the free energy of transfer per amino acid and
$\chi$ is the Flory-Huggins parameter, averaged over the amino acid
composition of the peptide and over the solvent accessibilities of
the various amino acids.  Following recent work \cite{Ghosh:2009},
we neglect the distinction between interior and surface residues
that was drawn in older models \cite{Dill:1985}.

\subsection{Modeling the protofilament and its nucleus}
The $\beta$-sandwich motif, shown in Fig. \ref{fig:key}c, is a
common feature of amyloid aggregates observed in NMR and X-ray
structures \cite{Nelson:2005,Luhrs:2005,Sawaya:2007,Wasmer:2008}.
Here, we assume that the basic structural element comprising
protofilaments and fibrils is the $\beta$-sandwich.  Before
describing our fibril model, we first define our terminology for
structures throughout this paper.  A \emph{$\beta$-strand} is a
single linear stretch of peptide chain. A \emph{$\beta$-sheet} is
comprised of two or more hydrogen-bonded \emph{$\beta$-strands} (see
Fig. \ref{fig:key}a,b). A \emph{$\beta$-sandwich} is two planar
$\beta$-sheets face-to-face (see Fig. \ref{fig:key}c).  For example,
in the fibrillar state the A$\beta$ molecule is a $V$-shaped
$\beta$-hairpin comprised of two beta strands.  In amyloid fibrils,
the $\beta$-sandwich is stabilized by H-bonds parallel to the fibril
axis and by hydrophobic and van der Waals interactions from the
interdigitation of side chains within the steric zipper between the
two $\beta$-strands (see Fig. \ref{fig:key}c) \cite{Nelson:2005}. At
a given stage of fibrillization, we assume a $\beta$-sheet is
composed of $h$ $\beta$-strands. Each $\beta$ strand contributes $b$
amino acids to the $\beta$-sheet. The sheet width $b$ must satisfy
$b\leq \ell $, where $\ell \leq L$ is the length of the
$\beta$-strands in the mature fibril. At a given stage of fibril
formation, the total number of residues in the $\beta$ state is $m=2
bh$, where the factor of two accounts for the two sheets in the
$\beta$-sandwich. The quantities $b$ and $h$ are shown schematically
in Fig. \ref{fig:sandwich}c,d.  The quantity $m$ serves as an order
parameter for the extent of fibril formation.

We treat the equilibrium between the oligomer (state B in Fig.
\ref{fig:sandwich}), the fibril nuclei (states C and D in Fig.
\ref{fig:sandwich}), the protofilament (E in Fig.
\ref{fig:sandwich}), and the full fibril (F in Fig.
\ref{fig:sandwich}) in a way that resembles the standard treatment
of the helix-coil transition in peptides
\cite{Zimm:1959,Schellman,LR,Poland:1970,JacsKings:2009}.

We call the states $BCDE$ the \emph{fibril ordering}
pathway.  The free energy is
\begin{equation}\label{eq:simpleorder}
\frac{\Delta F_{\mathrm{BCDE}}(m)}{kT} =-\chi (N L-m) - m \ln g_s -\sqrt{\frac{m}{2}}\ln
\gamma,
\end{equation}
as a function of $m$, which can be regarded as an `order parameter'
or a `reaction coordinate' along the route $BCDE$.  $m$ ranges from
$m=0$ when the system is fully disordered (\emph{i.e.,} fully in
state B), to $m=NL$ when the system is fully ordered in the
$\beta$-state (\emph{i.e.,} fully in state E).  (So, in normalized
form, a reaction coordinate could be expressed as $\xi = m/NL$). The
first term in Eq. \ref{eq:simpleorder} is the free energy of
converting $m$ of the $NL$ amino acids from their oligomeric
disordered state, with a corresponding loss of the disordered
micelle-like hydrophobic interactions.

The second term in Eq. \ref{eq:simpleorder}, $-m k T \ln g_s$, is
the free energy of forming $m$ pairwise steric zipper interactions
in the core of the fibril.  $g_s$ is a dimensionless propagation
equilibrium coefficient.  $g_s$ resembles the quantity $s$ in
helix-coil theories \cite{Zimm:1959,Poland:1970}, except that $g_s$
here describes $\beta$ structure, not $\alpha$-helical structure.
$g_s$ captures various types of interactions, including
conformational entropy, hydrogen bonds, steric packing, and ordered
sidechain hydrophobic interactions.  A necessary condition for
fibril formation is $g_s > 1$.  That is, fibrils can only form when
the sterically zipped state (hydrogen bonds, packing, and ordered
hydrophobic interactions) is more favorable than the monomeric
state.  The subscript $s$ here in $g_s$ indicates an interaction
within a single $\beta$-sandwich, not between the different
$\beta$-sandwiches that make up a full fibril.

We treat fibrillar ordering as a surface/interior nucleation
process.  The third term in Eq. \ref{eq:simpleorder}, $\sqrt{m/2}
\ln \gamma$, is the free energy of initiating steric zipping at the
edge of the $\beta$-sandwich (a square having $m/2$ residues,
has a perimeter with $\sqrt{m/2}$ residues).  In our model, $\gamma$, is a surface tension for forming the perimeter bonding.  In the metaphor of helix-coil theories, $\gamma$ resembles
$\sigma$, the helix-coil nucleation parameter. Fig.
\ref{fig:sandwich} shows that the edge of the $\beta$-sandwich has
$2b$ unsatisfied H-bonds, but only $b$ unsatisfied hydrophobic
contacts due to the stagger between the two sheets.  It is these
missing hydrophobic and H-bond interactions that account for why
there should be a barrier, $\gamma < 1$, to nucleating the fibril.
This nucleation barrier free energy in Eq. \ref{eq:simpleorder} is maximal for $b=\ell$. At the present stage of knowledge of microscopic structures, this square-shape approximation has the advantage of
simplicity and is adequate to capture the shift in the
oligomer-fibril transition from hydrophobic to combined hydrophobic
and hydrogen bonding interactions \cite{Auer:2008}.

This model gives insight into fibril formation rates.  We compute the free
energy of the transition state by finding the maximum value of $\Delta F_{\mathrm{order}}$ along the reaction coordinate $m$ using Equation~\ref{eq:simpleorder}.  The transition state is at $\Delta F^\ddagger = (d \Delta
F_{\mathrm{order}}/d m)_{m^\star}=0$, so
given by
\begin{equation}
\frac{\Delta \Delta F^\ddagger}{kT} =\frac{\Delta F_{\mathrm{order}}(m^\star)-\Delta
F_{\mathrm{order}}(0)}{kT} =\frac{\ln^2 \gamma}{8(\ln(g_s)-\chi)}.
\label{eq:barrierheight}
\end{equation}
Eq. \ref{eq:barrierheight} shows that the free
energy barrier will be extremely sensitive since the quantity $\ln(g_s)-\chi$ in the denominator will be small.  $\ln g_s$ and $\chi$ are dimensionless quantities of order unity; their difference is small because the zipping free energy is expected to be only slightly more favorable than amorphous hydrophobic interactions.  Small
variations in $g_s$ or $\chi$, at the level of single amino-acid
changes or slight changes in solution conditions, could change
fibrillization rates by several orders of magnitude
\cite{Jarrett:1993}.  This provides a rationale for understanding
how a single mutation could transform a normal fibrillization rate,
which might be too slow to cause disease on a human lifetime, into a
much faster fibrillization rate, sufficient to cause disease during
a human lifetime.

Our model of the amyloid nucleation process differs from classic
nucleation mechanisms in two respects: (1) ours involves a
one-dimensional line tension, rather than a two-dimensional surface
tension \cite{Muthu:2009}, and (2) our ordering transition is from
oligomers to fibrils, not from monomers to aggregates, so our
fibrillization mechanism is not driven by increasing the solution
concentration of monomers.  This is consistent with experiments
showing that amyloid nucleation is concentration-independent. The
proposed explanation in the Nucleated Conformational Conversion
(NCC) model \cite{Serio:2000} is that oligomeric chains must enter
an \emph{activated conformation} to proceed to fibrils.  In our
model, the role of activation is played by the entropically
unfavorable steric-zipper nucleus.

\subsection{Modeling the full fibril}

In our model, a full fibril consists of $p$ $\beta$-sandwich-motif
protofilaments stacked and bundled together.  Fig.
\ref{fig:sandwich}e shows such a fibril for $p=2$.  $\Delta
F_{\mathrm{EF}}$ is the free energy of bundling protofilaments into
fibrils,
\begin{equation}
\frac{\Delta F_{\mathrm{bundling}}}{kT} = \frac{\Delta
F_{\mathrm{EF}}}{kT} =  nL \epsilon = -nL\ln\left( \frac{g}{g_s}
\right) \label{eq:bundling}
\end{equation}
where L remains the number of amino acids per peptide chain.  $n$ is
the number of peptide chains in the bundled fibril.  $g$ is the
propagation constant for forming $\beta$ structure in the mature
fibril. Eq. \ref{eq:bundling} gives $-\ln g=-\ln g_s +\epsilon$,
so $-\ln g_s$ is the free energy of the interactions within a
single protofilament, and $\epsilon$ is the interaction energy holding the
protofilaments together.  Because these bundling interactions occur
only between a subset of residues on the perimeter of the
protofilament, we expect that $g$ and $g_s$ will be similar.

\begin{figure}[ht]
\vspace{0.6 cm}
\begin{center}
\includegraphics[angle=0,scale=0.9]{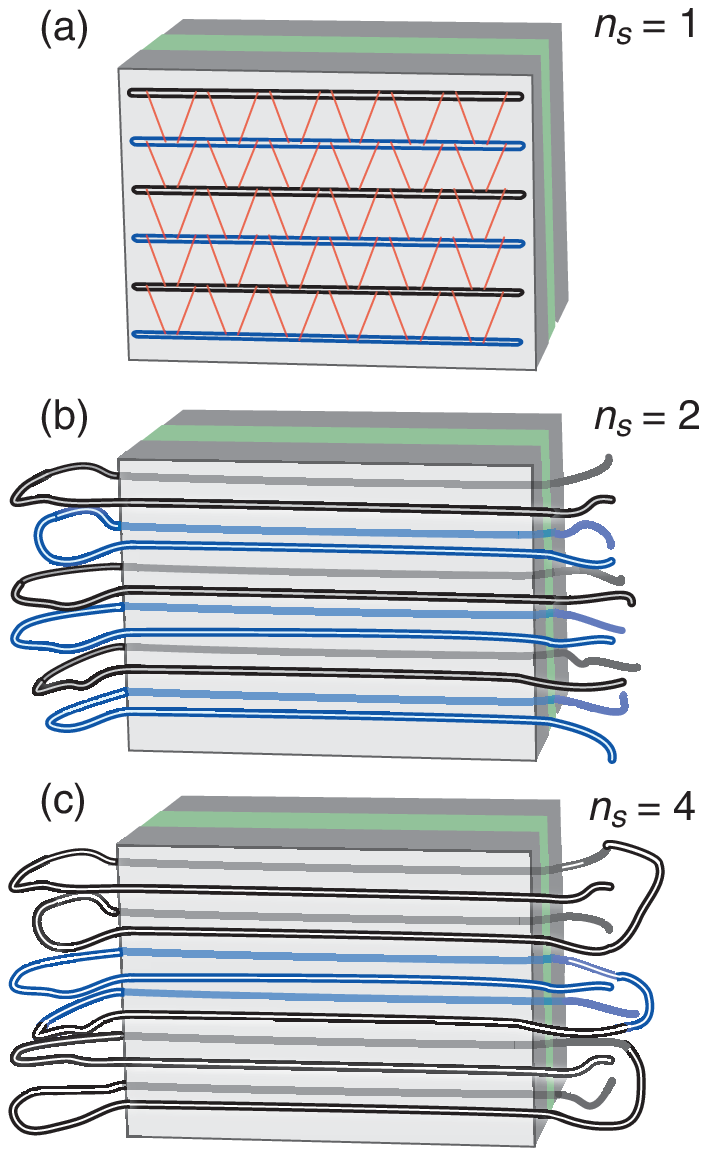}
\end{center}
\caption{ \label{fig:Nstrand} Schematic representation of the
parameter $n_s$.  Here each peptide chain contributes a) one, b)
two, and c) four $\beta$-strands to the fibril.  For clarity,
adjacent peptide chains are shown in alternating colors.}
\end{figure}

Peptides are frequently folded within a fibril such that a single
peptide chain contributes multiple $\beta$-strands to the fibril. To
account for this we introduce the parameters $\ell$, the length of
each $\beta$-strand, and $n_s$, the number of $\beta$ strands formed
by each peptide chain.  For example, peptides such as A$\beta$ and
IAPP that form a single hairpin in the mature fibril have $n_s=2$.
The quantity $n_s$, defined such that $L=n_s \ell$ is shown in Fig.
\ref{fig:Nstrand}.

From Eq. \ref{eq:simpleorder} it is clear that each protofilament
will incur a nucleation penalty $-kT \ell \ln \gamma$ and each
$\beta$-strand in the fibril will contribute a binding energy
$-kT\ell\ln g$. However, a more convenient quantity is the binding
energy per peptide $-kT\ell n_s\ln g=-kTL\ln g$. The free energy of
a fibril consisting of $n$ peptides is then
\begin{equation}\label{fibrilF}
\frac{\Delta F_n}{kT} =\frac{\Delta
F_{\mathrm{AF}}}{kT}=-\ln(\gamma^{p\ell}g^{n L}).
\end{equation}
The fibrillization index $n$ must be greater than the minimum fibril
size $n_0$, however we will find that the physical observables are
insensitive to the precise value of $n_0$. This is in contrast to
the oligomer size $N$ which plays an important role in the phase
behavior.

\subsection{The monomer-oligomer-fibril assembly equilibrium}

Now, we combine the free energies above into a grand canonical
ensemble to determine how the assembly equilibria depend on the
concentration of peptide monomers in solution.  If the oligomeric
state resembles a micelle, a reasonable approximation is that the
oligomer species is dominated by a single aggregation number, with
free energy given by Eq. \ref{eq:Foligo}.  However, for the fibril,
we assume a continuum of aggregation states having number $n$
peptide chains and a free energy given by Eq. \ref{fibrilF}. To
compute the properties of the solution, we compute the \emph{binding
polynomial} \cite{Dill:MDF},
\begin{equation}
Q=c_1+ c_1^N e^{\chi N L}+\gamma^{p\ell} \sum_n c_1^n g^{n L},
\label{eq:bindingpoly}
\end{equation}
In Eq. \ref{eq:bindingpoly}, $c_1$ is the concentration of monomers,
$c_1^N e^{\chi N L}$ is the concentration of oligomers, and the
final term is the sum over the concentrations of the fibrils of all
possible lengths.  The total peptide concentration, $c_0$ can be
computed by using the concentration of each species and summing the
number of peptides in each species. Thus, $c_0$ is written as
\begin{equation}
c_0 = \frac{d Q}{d \ln c_1} = c_1 + N c_1^N e^{\chi N L} + \gamma^{p
\ell } \sum_{n=n_0}^{\infty} n c_1^n g^{n L}, \label{eq:bulkwithsum}
\end{equation}
where $n_0$ is the smallest accessible fibril size.

The solution phase behavior is given by the peptide concentrations
in each of three states: monomer, $c_1$; oligomer,
$c_{\mathrm{oligo}}$; and fibril, $c_{\mathrm{fibril}}$, where
\begin{equation}\label{eq:coligo}
c_{\mathrm{oligo}} = N c_1^N e^{\chi N L}
\end{equation}
and
\begin{equation}\label{eq:cfibril}
c_{\mathrm{fibril}} = \gamma^{p \ell } \sum_{n=n_0}^{\infty} n c_1^n
g^{n L}
\end{equation}
are the component terms from Eq.~\ref{eq:bulkwithsum}.  The three
relative concentration quantities, $c_1/c_0, c_{\mathrm{oligo}}/c_0$
and $c_{\mathrm{fibril}}/c_0$ must sum to one. To compute the phase
diagram we numerically solve Eq. \ref{eq:bulkwithsum} for $c_1$ at
fixed values of $c_0$, $g$, $\gamma$, and $\chi$.  The
concentrations of peptides in the fibril and oligomer states are
then computed from Eqs. \ref{eq:coligo} and \ref{eq:cfibril}. Fig.
\ref{fig:phasevarchi} shows the computed phase diagrams.  The
boundaries in Fig. \ref{fig:phasevarchi} represent the conditions of
equal populations of the two corresponding states.  In the Methods
section we derive analytic expressions for the phase boundaries.
These are shown by the black lines in Fig. \ref{fig:phasevarchi}.
The model predictions are given in the following section.

\section{Results}

\begin{figure}[ht]
\vspace{0.6 cm}
\begin{center}
\includegraphics[angle=0,scale=0.9]{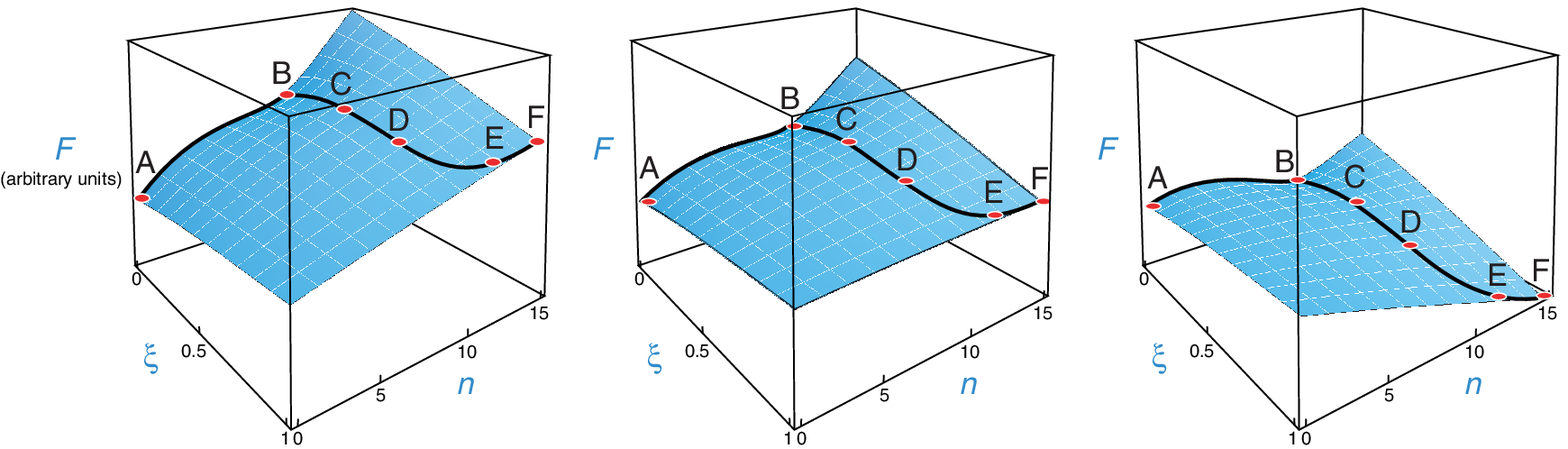}
\end{center}
\caption{ \label{fig:landscape} Schematic representation of the free
energy landscape described by our model at low (left), intermediate
(middle), and high peptide concentration (right).  Labels correspond
to the states shown in Fig. \ref{fig:sandwich}.  At low
concentrations the monomer state (A) is the free energy minimum,
while at high concentrations the fibril (F) is the minimum.  At
intermediate concentrations the solution is an equilibrium of
monomers and oligomers (B).}
\end{figure}

The model defined in the previous section leads to a free energy
landscape with features schematically shown in Fig.
\ref{fig:landscape}.  In the following we compute the phase
equilibria resulting from this landscape.

\subsection{The model predicts an \emph{amyloid triple point}, a 3-state equilibria}

\begin{figure}[ht]
\vspace{0.6 cm}
\begin{center}
\includegraphics[angle=0,scale=1.0]{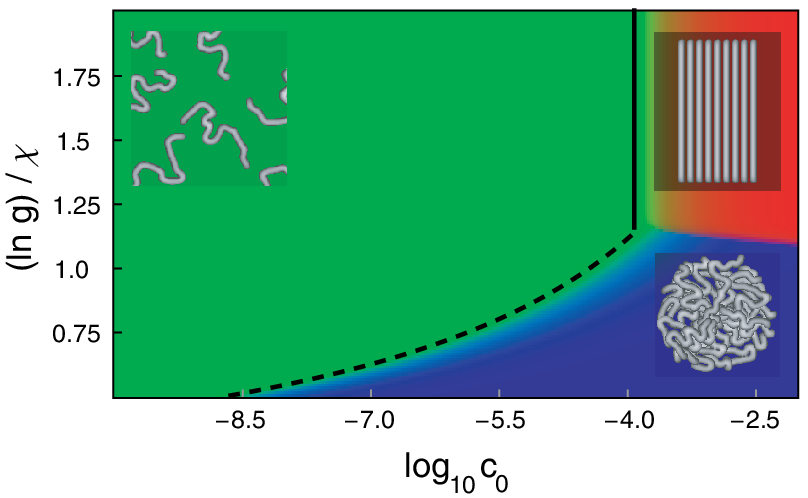}
\end{center}
\caption{ \label{fig:phasevarchi} Phase diagram for peptides with
$p,n_s=1$, $\ell =15$, $\ln g=0.6$, and $\ln \gamma=-2$ as a
function of peptide concentration and $\ln g/\chi$.  Green
=$c_1/c_0$ (monomers).  Blue = $c_{\mathrm{oligo}}/c_0)$
(oligomers).  Red =$c_{\mathrm{fibril}}/c_0$ (fibrils).  Lines
depict phase boundaries computed from Eq.
\ref{eq:fibriloligoboundary} (long dashes), Eq. \ref{eq:COCapdx}
(short dashes), and Eq. \ref{eq:CFCapdx} (solid).}
\end{figure}

Fig. \ref{fig:phasevarchi} shows the phase diagram computed from Eq.
\ref{eq:bulkwithsum}.  The $x$-axis shows the monomer concentration.
The $y$-axis shows $\ln g/\chi$, the ratio of the free energy for a
steric zipper to the free energy for amorphous hydrophobic
aggregation. The model predicts three main features.  First,
increasing the amyloid peptide concentration in solution leads to
increased aggregation (both oligomers and fibrils).  Second, not
surprisingly, at high peptide concentrations, changing solution
conditions to favor steric zipping tips the balance from oligomers
toward fibrils. This phase equilibrium line is relatively flat,
indicating that it is not very dependent on monomer concentration.
Third, there should be a triple point, a particular monomer
concentration and solution condition at which monomers, oligomers,
and fibrils are all present in equal populations.

The phase diagram can be closely approximated by comparing the
critical fibril concentration (CFC) to the critical oligomer
concentration (COC) (see Methods). The lesser of these two
quantities determines the aggregate species that appears upon
raising the peptide concentration. However, if the COC is less than
the CFC it may be possible to drive the solution from the oligomer
state to the fibril state by further raising the peptide
concentration.  This transition may be computed using the
fibril-oligomer coexistence condition (see Eqs.
\ref{eq:fibriloligoboundary} and \ref{eq:ureaFOboundary}). The
converse is not true; it is not possible to reach the oligomer phase
from the fibril phase by increasing the peptide concentration.  This
asymmetry arises from the definitions of the critical
concentrations.  The CFC is defined by the radius of convergence of
Eq. \ref{eq:bulkwithsum}, and thus it sets a hard limit on the
achievable monomer concentration.  On the other hand, the monomer
concentration will still rise, albeit weakly, upon reaching the COC,
and therefore it is possible for the monomer concentration to reach
the CFC even after oligomers have begun to form provided the CFC is
not much greater than the COC.

\subsection{When fibrils are stable, oligomers are not.}

Interestingly, the model predicts a `sponge-like behavior': under
fibril-forming conditions, amyloid peptide will be `soaked up' into
the fibrils and depleted from the oligomers.  To see this,
substitute the CFC, Eq. \ref{eq:CFCapdx}, into Eq. \ref{eq:coligo},
to get
\begin{equation}
c_{\mathrm{oligo}}\sim N e^{-LN(\ln g-\chi)}.
\label{eq:ColigoFibril2}
\end{equation}
This small quantity, $e^{-(\ln g-\chi)}<1$, is raised to a large
power, $LN$. So, unless $\chi$ and $\ln g$ are closely matched, the
concentration of the oligomeric state will be negligible under
fibril-forming conditions.  The implication for disease is that if
oligomers are toxic, promoting fibril formation may deplete the
toxins.

\subsection{Fibril concentration increases as a nonlinear function of monomer concentration.}

\begin{figure}[ht]
\vspace{0.6 cm}
\begin{center}
\includegraphics[angle=0,scale=1.0]{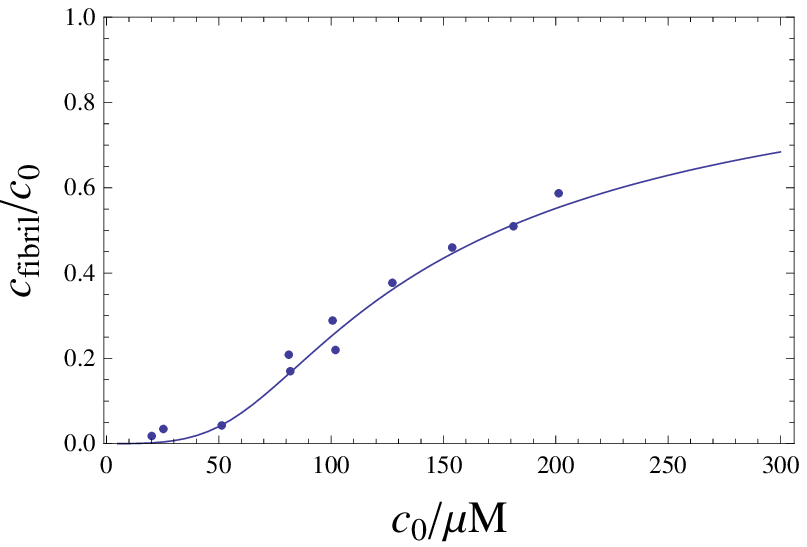}
\end{center}
\caption{ \label{fig:CDfit} Plot of $c_{fibril}$ as a function of
the bulk peptide concentration compared to CD data of Terzi {\em et
al.} for A$\beta_{1-40}$ \cite{Terzi:1995}. $L =26$, $n_s=2$, $p=2$,
$g=1.66$ and $\gamma=0.54$}
\end{figure}

Fig. \ref{fig:CDfit} compares the theory for how the fibril
population, $c_{\mathrm{fibril}}$, depends on peptide concentration,
to the experiments of Terzi et al. \cite{Terzi:1995}.  Since the
N-terminal 11-16 residues of A$\beta$ are disordered
\cite{Petkova:2002,Luhrs:2005} in the fibril state, we take $L =26$.
In order to fit the data we convert the experimental concentration
$c_M$ (in Molar units) to the dimensionless concentrations required
in our treatment.  We use $c=(c_M/55.5) M$. From that fit, we find
that $-L \ln g=-13.1$ and $-p \ell \ln \gamma =15.5$ at the
experimental conditions of 278 K and pH 7.4. This is an order of
magnitude stronger than the per-residue binding energy within native
proteins \cite{Ghosh:2009}.

\subsection{Fibril lengths undergo a `growth transition' \emph{vs.} monomer concentration.}

Now we compute the distribution of fibril lengths.  The
probability $P(n)$ that a fibril has a length $n$ is given by
\begin{equation}
P(n)=\frac{c_1^n g^{n L}}{\sum_i c_1^i g^{i L}}. \label{eq:Plength}
\end{equation}

\begin{figure}[ht]
\vspace{0.6 cm}
\begin{center}
\includegraphics[angle=0,scale=1.0]{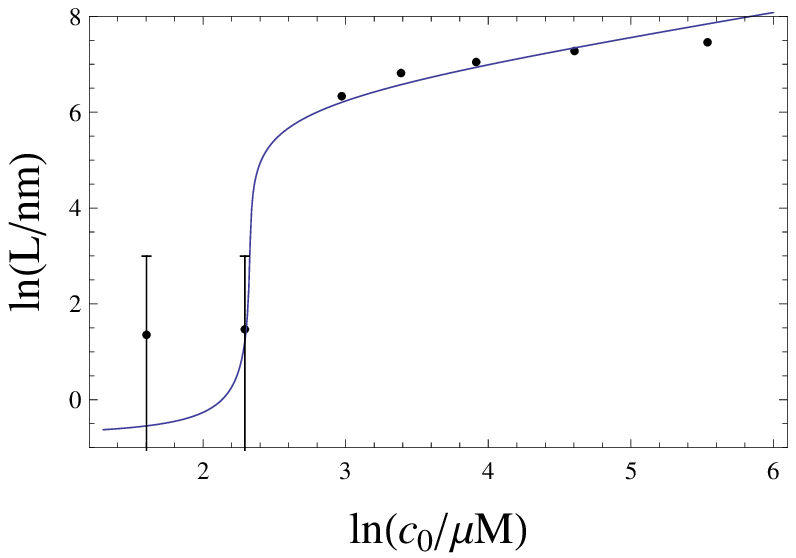}
\end{center}
\caption{ \label{fig:lengthfit} Average length of fibrils \emph{vs.}
peptide concentration, and compared to experiments on
$\alpha$-synuclein \cite{vanRaaij:2008}. }
\end{figure}

What are the average fibril lengths?  In the Methods section we show
that the average length scales as $c_0^{1/2}\gamma^{-{\ell p}/2}$,
in agreement with the concentration dependence found in reference
\cite{Lee:2009}.  Fig. \ref{fig:lengthfit} shows the predictions of
Eq. \ref{eq:lengthsums} compared to the average fibril-length
measurements of van Raaij et al \cite{vanRaaij:2008}. From the fit
we find, $g^{L}$, which determines the onset of fibrillization and
we find $\gamma^{\ell p}$, which determines the fibril length. We
obtain $-L \ln g=-15.5$ and $-\ell p \ln \gamma= 32.3$. In
$\alpha$-synuclein fibrils, it is found that $p=4$, twice the value
of A$\beta$ fibrils \cite{vanRaaij:2008}. Accounting for this factor
of two shows that the values of $L\ln g$ and $\ell  \ln \gamma$ are
quite similar to those determined for A$\beta$ in the previous
section.  $n_s$ and $\ell $ are not yet known for
$\alpha$-synuclein.

\begin{figure}[ht]
\vspace{0.6 cm}
\begin{center}
\includegraphics[angle=0,scale=0.7]{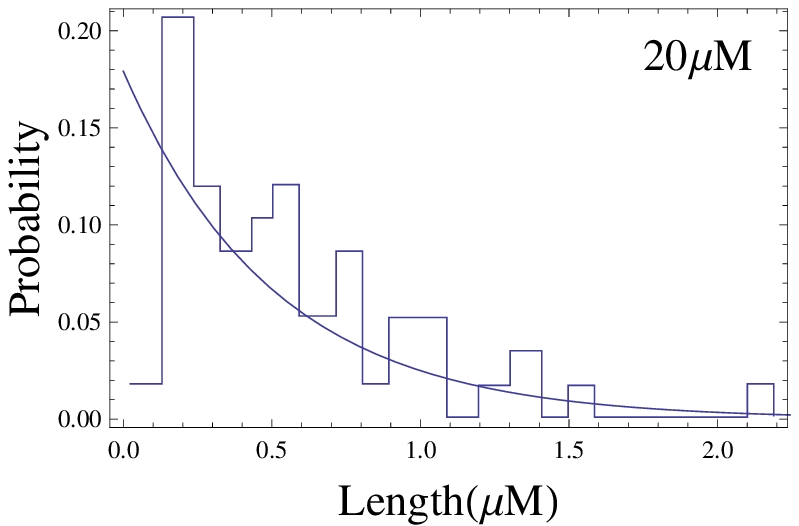}
\includegraphics[angle=0,scale=0.7]{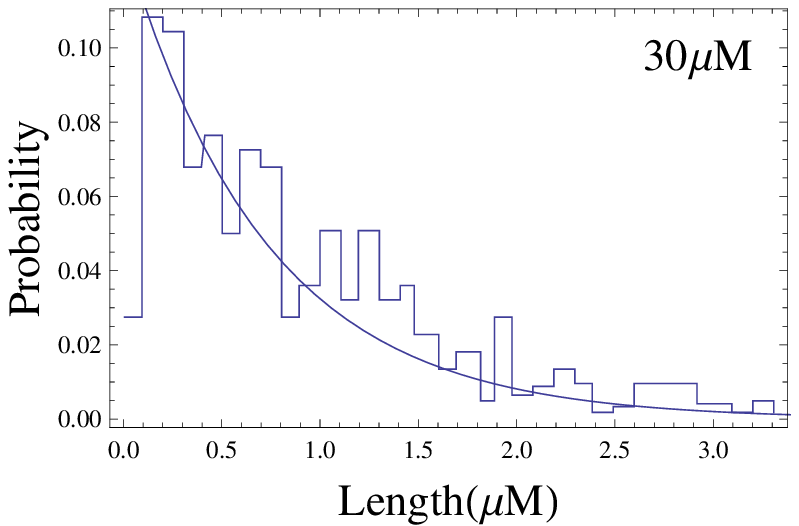}
\includegraphics[angle=0,scale=0.7]{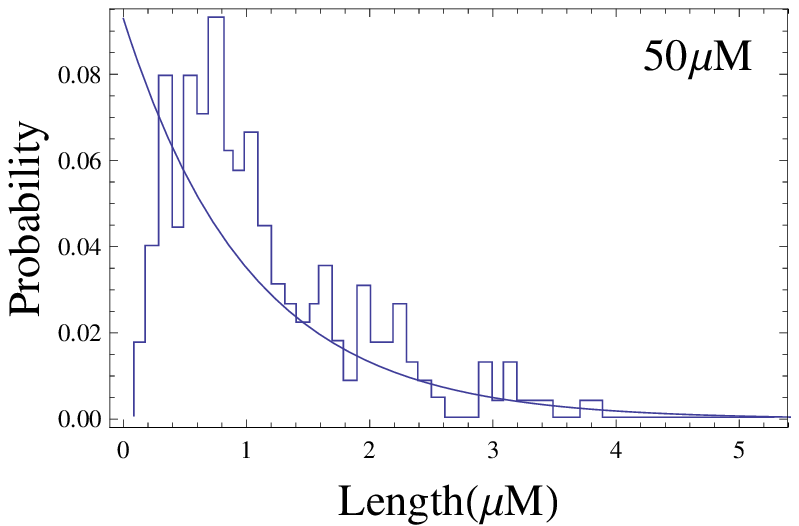}
\includegraphics[angle=0,scale=0.7]{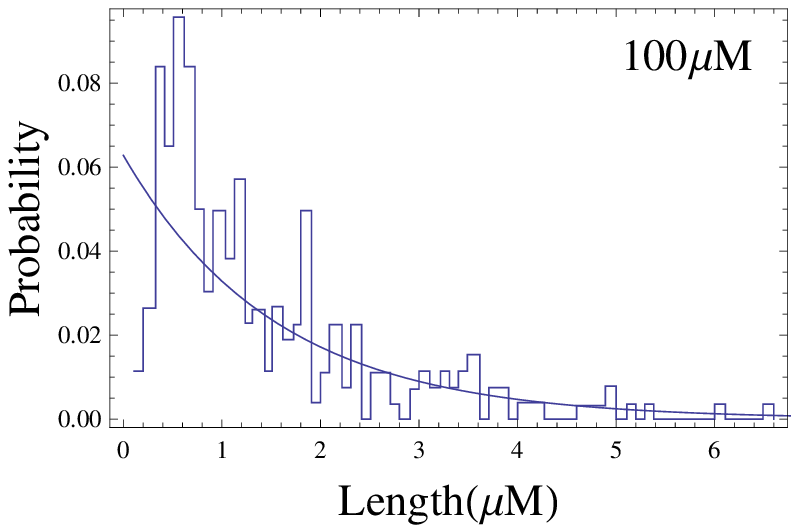}
\includegraphics[angle=0,scale=0.7]{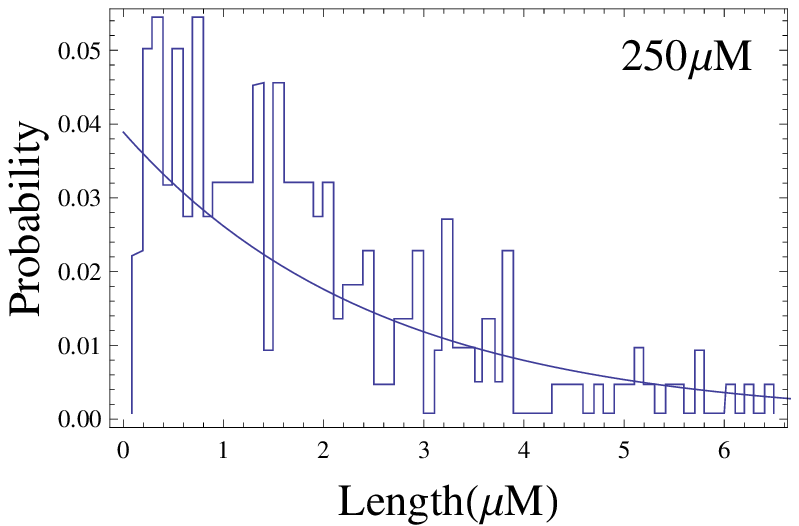}
\end{center}
\caption{ \label{fig:lengthdist} Comparison of the computed fibril
length distribution as a function of the bulk peptide concentration
to the experimental distributions of $\alpha$-synuclein
\cite{vanRaaij:2008}.}
\end{figure}

Fig. \ref{fig:lengthdist} shows the prediction that the fibril lengths follow an exponential distribution; see Eq. \ref{eq:lengthsums}.

\subsection{Denaturants destabilize the fibrils and oligomers}
\begin{figure}[ht]
\vspace{0.6 cm}
\begin{center}
\includegraphics[angle=0,scale=1.0]{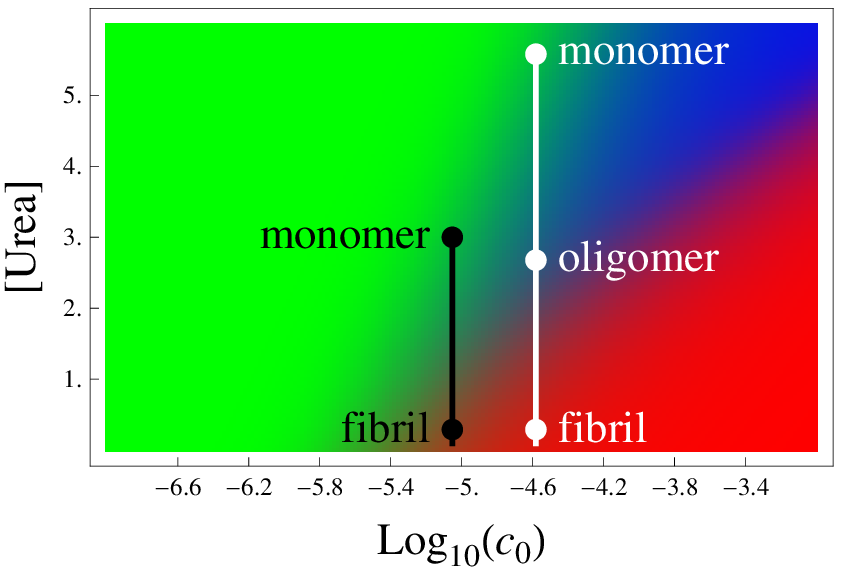}
\end{center}
\caption{ \label{fig:phaseurea} Phase diagram for A$\beta$ as a
function of peptide concentration and urea concentration.  The
colors represent monomers (green), oligomers (blue), and fibrils
(red).   $N=4$, $\chi NL=36.4$ \cite{Chen:2006}, and all other
parameters are identical to Fig. \ref{fig:CDfit}.  This diagram
explains a discrepancy between the experiments of Chen and Glabe, where the black line indicates denaturation.  They found no intermediate oligomers.  For the Kim experiments, denaturation is indicated by the white line; they
observed oligomeric intermediates.}
\end{figure}

What is the effect of denaturants and osmolytes on amyloid
aggregation?  First, because both oligomers and fibrils are stabilized by hydrogen bonding and hydrophobic interactions, denaturants should `melt out' amyloid aggregated states.  A more subtle question is how denaturants shift the oligomer-fibril equilibrium.  Fig. \ref{fig:phaseurea} shows the model
predictions (1) that denaturants such as urea, not unexpectedly,
should weaken hydrophobic and hydrogen bonding interactions,
disrupting aggregation, (2) that more denaturant is required to
disrupt aggregates if the amyloid concentration is high, and (3)
that adding denaturant to fibrils can drive the system into the
oligomer state.

We note two additional points.  First, in apparent contradiction to this prediction, denaturants are sometimes used to promote fibrillization, but that appears to be observed
exclusively in systems having a native folded state
\cite{Hamada:2002,Ahmad:2004,Wang:2007}, unlike the systems we model
here.  Second, the present model resolves a paradoxical result in
the literature.   Chen and Glabe found that urea drove fibrils to
`melt' directly to monomers without passing through the oligomer
state \cite{Chen:2006}.  In contrast, Kim \emph{et al.} found that urea
drove fibrils to melt to oligomers, when then melted to monomers
\cite{Kim:2004}. Our Fig. \ref{fig:phaseurea} gives an explanation:
the A$\beta$ peptide concentrations used by Kim \emph{et al.} were 3-10-fold greater than those used by Chen and Glabe, shifting to a region of the phase diagram in which oligomers are a stable intermediate
phase.

\subsection{Electrostatic repulsion destabilizes the fibrils}

To treat the effects of pH and salt, we express the binding
free energy
\begin{equation}
-L \ln g(q,c_s)=-L \ln g_0 +\Delta F_{es}(q,c_s)
\end{equation}
in terms of $g_0$, which accounts for the binding energy for a
reference peptide having zero net charge, and an electrostatic
component, $\Delta F_{es}$, which is the free energy of charging up
the peptides from their uncharged state to a net charge $q$ in the
presence of a salt concentration $c_s$. The latter is given
quantitatively by Eq \ref{eq:Fes}.

\begin{figure}[ht]
\vspace{0.6 cm}
\begin{center}
\includegraphics[angle=0,scale=0.9]{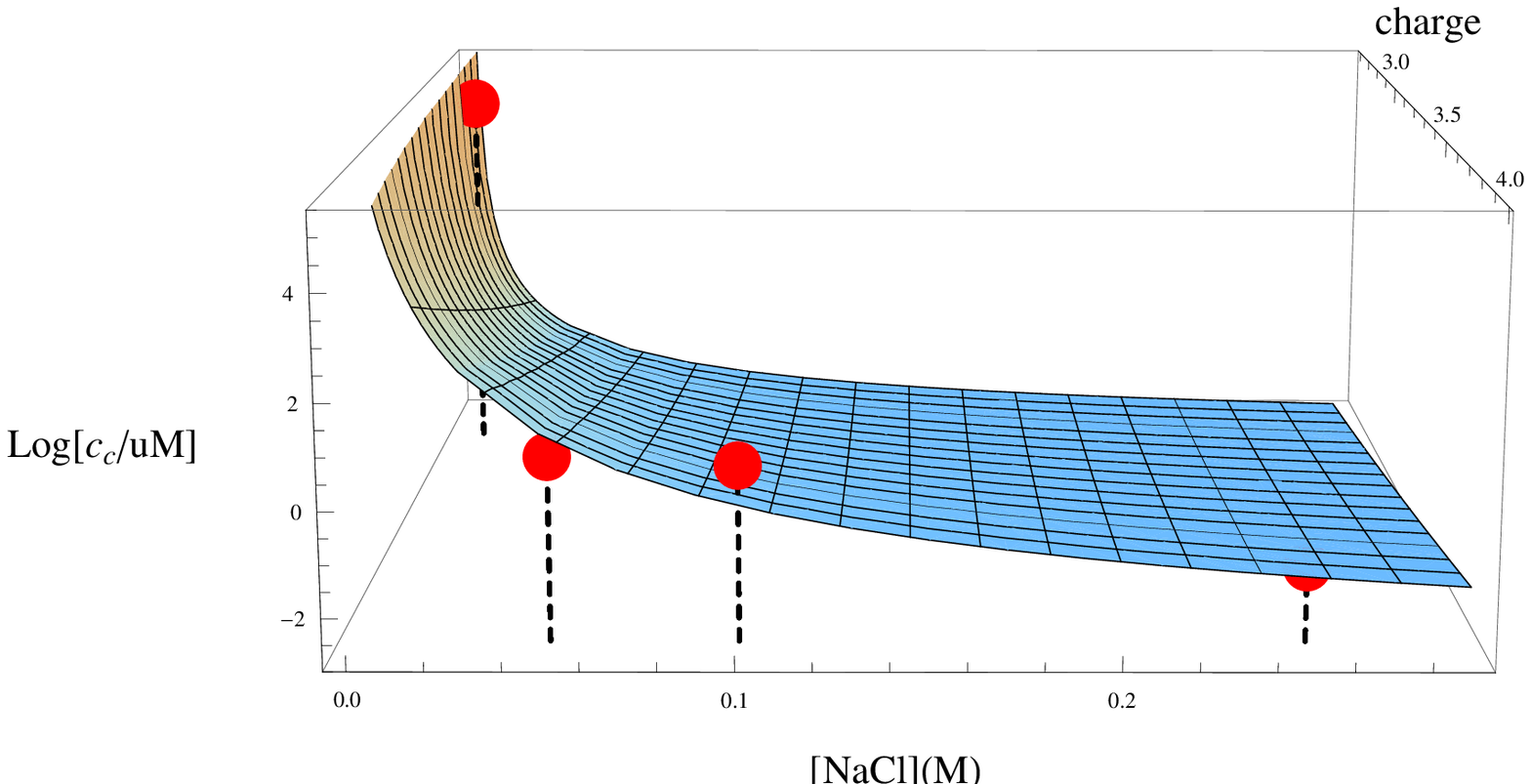}
\end{center}
\caption{ \label{fig:salt} Predicted solubility of A$\beta_{40}$ as a function of salt concentration and net peptide charge. Data points at $q=3.9$ are from \cite{Klement:2007} and the point at $q=2.8$ is from \cite{Terzi:1995}.}
\end{figure}

Fig. \ref{fig:salt} compares a simple calculation of the
electrostatic repulsions with experimental measurements of
dependence of the critical concentration, which we identify with the
theoretical quantity $c_1^{(f)}$, on salt concentration.  We compute
a charge of $q=-3.9$ at pH 9.0 and $q=-2.8$ at pH 7.4 for the
A$\beta$ peptide \cite{Sillero:1989}.  We treat each peptide as a
cylinder of radius $R=10$nm \cite{Klement:2007} using
Poisson-Boltzmann theory; see Eq. \ref{eq:Fes}.  Fig. \ref{fig:salt}
shows two experiments: Klement et al. for pH 9 and Terzi et al. for
pH 7.4.  The sole fitting parameter in Fig. \ref{fig:salt} is $-L\ln
g_0=-21.9$.  This corresponds to $-L\ln g=-15.5$ under the pH 7.4,
5mM conditions of Terzi et al. This represents a good agreement with
our previous result of $-L\ln g=-13.1$ given the extreme sensitivity
of the electrostatic free energy at such low salt concentrations.

\section{Conclusion}

We have described a general thermodynamic theory for the aggregation
of short peptides into globular and fibrillar aggregates.  The model
predicts two transitions: (1) a micelle-like transition of monomeric
peptides in solution to an oligomeric state involving a loose
hydrophobic core and a loss of translational entropy, and (2) a
transition from the disordered globular oligomeric state to an
ordered $\beta$-structured fibrillar state, driven by tighter
packing, hydrogen bonding, and steric and hydrophobic interactions.

We find good agreement of the model with experiments on fibril
concentrations, average fibril lengths, and fibril length
distributions \emph{vs.} monomer concentrations.  We find that the
phase boundaries and transition states are highly sensitive to small
changes in solution conditions and protein properties.  Such
sensitivities may be relevant to aggregation processes in amyloid
diseases.


\section{Methods}

{\bf Critical fibril concentration (CFC)}

The concentration of peptides in the fibril state is given by the
final term in Eq. \ref{eq:bulkwithsum}
\begin{eqnarray}
c_{fibril}&=&\gamma^{\ell p}\sum_{n=n_0}^\infty n
c_1^n g^{n L} \label{eq:cfibrilsum}\\
&=&\gamma^{\ell p}c_1 g^{L}\frac{d}{d(c_1 g^{L})}\sum_{n=n_0}^\infty
c_1^n g^{n L} \\
&=& \gamma^{\ell p}\frac{n_0(1-c_1 g^{L})(c_1
g^{n_s\ell})^{n_0}+(c_1 g^{L})^{n_0+1}}{(1-c_1 g^{L})^2}.
\end{eqnarray}
The sum in Eq. \ref{eq:cfibrilsum} converges when the argument is
less than unity, therefore the monomer concentration must satisfy
$c_1<g^{-L}$ if the bulk concentration is to remain finite. This
radius of convergence defines a concentration
\begin{equation}
c_1^{(f)}\sim g^{-L}, \label{eq:CFCapdx}
\end{equation}
which may be interpreted as the CFC for the fibril solution.  For
$c_1$ much less than this value the concentration of monomers in the
fibril state $c_{fibril}$ is strongly suppressed by the factor
$\gamma^{\ell p} (c_1 g^{L})^{n_0}$. However, as $c_1$ approaches
$g^{-L}$ the fibril concentration diverges. In this regime we can
write $c_1 g^{L}=1-\delta$ and the fibril concentration becomes
\begin{eqnarray}
c_{fibril}&\simeq&\gamma^{ \ell p} \frac{(1-n_0 \delta)(1+n_0\delta -\delta)}{\delta^2}\\
&=&\gamma^{ \ell p} \frac{c_1 g^{L}+{\cal O}(\delta^2)}{(1-c_1
g^{L})^2} \label{eq:Cfibril},
\end{eqnarray}
which demonstrates that $c_{fibril}$ is insensitive to the lower
limit $n_0$.

Note that our definition for the CFC differs from that given in reference
\cite{Lee:2009}.

{\bf Critical oligomer concentration (COC)}

Due to the importance of oligomers in disease progression, we would
like to calculate the concentration of oligomers both in the
presence and absence of fibrils.  In the presence of fibrils we can
approximate the concentration of peptides in the oligomer state,
$c_{oligo}$, by using the CFC in the second term of Eq.
\ref{eq:bulkwithsum} since the monomer concentration varies little in
the vicinity of the fibril CFC
\begin{equation}
c_{oligo}\sim N g^{-LN}e^{\chi LN}. \label{eq:ColigoFibril}
\end{equation}
This expression is only valid provided $e^{\chi}<g$.  If this
condition is not satisfied, then Eq. \ref{eq:simpleorder} is a
monotonically increasing function of $m$ meaning the oligomer state
has a lower free energy per peptide than a fibril of any length.  In
this case no fibrils will be formed, and the solution will be an
equilibrium mixture of oligomers and monomers.  We define the COC to be the
concentration where the oligomer and monomer states have equal
occupancies.  Using the appropriate terms from Eq.
\ref{eq:bulkwithsum} we find the COC given by
\begin{equation}
c_1^{(COC)} = \left(\frac{e^{-\chi NL}}{N}
\right)^{1/(N-1)}. \label{eq:COCapdx}
\end{equation}

{\bf Fibril-oligomer boundary}

The boundary between the fibril and oligomer phases is defined, for
points suitably removed from the monomer phase, by the condition
$c_{fibril}=c_{oligo}\simeq c_0/2$. Using Eqs. \ref{eq:Cfibril} and
\ref{eq:ColigoFibril} we have
\begin{eqnarray}
c_0/2&=&\gamma^{ \ell p} \frac{c_1 g^{L}}{(1-c_1 g^{L})^2} \label{eq:fibrilboundary}\\
c_0/2&=&Nc_1^N e^{\chi NL}.\label{eq:oligoboundary}
\end{eqnarray}
Eq. \ref{eq:fibrilboundary} yields a recursive formula for $c_1$,
which to lowest order gives
\begin{equation}
c_1 \simeq g^{-L}(1-\sqrt{2\gamma^{\ell p}/c_0})
\end{equation}
which can be combined with Eq. \ref{eq:oligoboundary} to yield a
condition for the phase boundary
\begin{equation}
\frac{\ln g}{\chi}=\frac{NL\ln g}{\ln\left(\frac{c_0}{2N g^{-L}}
\right)-N\ln\left(1-\sqrt{2\gamma^{\ell p}/c_0}\right)}.
\label{eq:fibriloligoboundary}
\end{equation}
This expression is plotted with long dashes in Fig.
\ref{fig:phasevarchi}

{\bf Average fibril length}

The critical concentrations for fibril and oligomer formation,
$c_1^{(f)}$ and $c_1^{(o)}$ are notably lacking a dependence on the
nucleation parameter $\gamma$.  While this parameter has little
effect on the relative stability of the fibril and oligomer phases,
we expect that it will play a large role in determining the
equilibrium lengths of mature fibrils.  To see this we consider a
system that is deep within the regime where fibrils are the dominant
species so that $c_0\simeq c_{fibril}$. Using Eq. \ref{eq:Cfibril}
we find
\begin{equation}
c_1\simeq g^{-L}(1-\sqrt{\gamma^{\ell p}/c_0}).
\label{eq:onlyfibrils}
\end{equation}
The $j$th moment of the fibril length distribution is given by
\begin{eqnarray}
\langle n^j \rangle &=& \frac{\gamma^{\ell p} \sum_n n^j c_1^n g^{n
L}}{\gamma^{\ell p} \sum_n c_1^n g^{n L}}
\label{eq:lengthsums} \\
&=&\left( \frac{(c_1 g^{L})^{n_0}}{1-c_1 g^{L}}\right)^{-1}
\left(c_1 g^{L}\frac{d}{d(c_1 g^{L})}\right)^j\left( \frac{(c_1
g^{L})^{n_0}}{1-c_1 g^{L}}\right). \label{eq:lenghtmoments}
\end{eqnarray}
The average length is given by the first moment $j=1$
\begin{eqnarray}
\langle \ell \rangle &=& n_0 +\frac{c_1 g^{L}}{1-c_1 g^{L}} \\
&\sim& c_0^{1/2}\gamma^{-{\ell p}/2} +
const\label{eq:lengthscaling},
\end{eqnarray}
where Eq. \ref{eq:onlyfibrils} has been used to extract the scaling
behavior.

The lack of a dependence on $g$ in this scaling relation is a result
of an approximation based on $c_0 \gg c_1^{(f)}$. When this
assumption is satisfied the large majority of protein is in the
fibril state, and the problem of determining the fibril lengths is
reduced to a question of fibril breakage statistics. Since each
breakage incurs a statistical penalty $\gamma^{\ell p}$, the
functional form of  Eq. \ref{eq:lengthscaling} is not surprising
(the factor of two in the exponent is a result of the degeneracy of
breakage points).  For systems near the onset of fibrillization $c_0
\simeq c_1^{(f)}$ the fibril lengths depend sensitively on $g$ and
Eq. \ref{eq:lengthsums} must be used to model the lengths.

{\bf Denaturants}

Denaturants destabilize the folded states of proteins by weakening
hydrophobic interactions and peptide-peptide H-bonds relative to
peptide-solvent H-bonds \cite{Auton:2007}.  To capture the effect of
denaturants on the interaction free energy quantities $g$ and
$\gamma$, we use
\begin{eqnarray}kT\ln g_c &= & kT(\ln g - a_0 c_d) \nonumber \\
kT\ln \gamma_c & = & kT(\ln \gamma + a_0 c_d), \label{eq:denatg}
\end{eqnarray}
where $g_c$ and $\gamma_c$ are the propagation parameters in the
presence of denaturant, $c_d$ is the denaturant concentration and
$a_0$ is a constant describing the destabilizing effect of the
osmolyte. This form reflects the fact that the denaturant weakens
the H-bonds captured in $g$, but also reduces the fibril end free
energy $-\ln \gamma$, which arises largely from unsatisfied H-bonds.

Using Eqs. \ref{eq:CFCapdx}, \ref{eq:COCapdx}, and \ref{eq:denatg}
we can compute how the critical concentrations will shift as a
function of denaturant concentration.  At the onset of
fibrillization we have $c_0=c_1^{(f)}$, so from Eqs.
\ref{eq:CFCapdx} and \ref{eq:denatg} we have
\begin{eqnarray}
\ln c_1&=&-L \ln g_c \\
&=&-L(\ln g -a_0 c_d),
\end{eqnarray}
which can be solved for $c_d$ to give the phase boundary
\begin{equation}
c_d=\frac{\ln c_0+L \ln g}{L a_0}.
\end{equation}
Similarly, for the oligomer state we write
\begin{equation}
\chi_c=\chi-a_1 c_d
\end{equation}
where $a_1$ has been introduced to reflect the fact that since the
oligomers are more dependent on hydrophobic interactions and less
dependent on H-bonds for stability, and therefore, the
destabilization coefficient will, in general, be different.   The
onset of oligomerization may be determined from Eq. \ref{eq:COCapdx}
\begin{equation}
c_0=\left(\frac{e^{-NL(\chi-a_1
c_d)}}{N}\right)^{1/(N-1)},
\end{equation}
which can be rearranged to yield
\begin{equation}
c_d=\frac{1}{a_1}\left(\chi+\frac{(N-1)\ln c_0 + \ln N}{NL}\right).
\end{equation}

The fibril-oligomer boundary can be derived from Eqs.
\ref{eq:fibrilboundary} and \ref{eq:oligoboundary}
\begin{equation}
\ln c_0/2N=NL(\chi - a_1 c_d)-NL(\ln g -a_0 c_d),
\label{eq:fiboligodenat}
\end{equation}
where we have dropped the correction term in Eq.
\ref{eq:fibrilboundary}.  In Fig. \ref{fig:phaseurea} we take
$a_1=a_0/2$ reflecting our expectation that the hydrophobic
interactions stabilizing the oligomer are less affected by the
presence of denaturant than the H-bonds stabilizing the fibril
\cite{Bolen1,Ghosh:2009}.  However, the particular choice of $a_0/2$
is for illustration purposes.  With this approximation for $a_1$ Eq.
\ref{eq:fiboligodenat} becomes
\begin{equation}
c_d=\frac{2}{a_0 NL}\left(\ln\frac{c_0}{2N}-NL(\chi+\ln g)\right).
\label{eq:ureaFOboundary}
\end{equation}

Based on denaturation studies, we expect that $a_0=0.022 M^{-1}$ for
urea and 0.042 for guanidinium \cite{Ghosh:2009}. Using our
estimates of $L \ln g = 13.1$ and $\ell p \ln \gamma = -15.5$ (in
the absence of urea) for A$\beta$ from our previous analysis of
fraction fibril as a function of concentration (Fig.
~\ref{fig:CDfit}) we can predict the fibril fraction as a function
of urea using Eq. ~\ref{eq:denatg} without a fit parameter. Our
prediction is compared with the experimental data in
Table~\ref{tab:ureatable}.

\begin{table}
\begin{tabular}{|c||c|c|c|}
\hline
&\multicolumn{2}{c|}{Experiment \cite{Kim:2004}} & Theory \\
\hline  [urea]& Oligomer fraction & Fibril fraction & Fibril
fraction \\
\hline 0.4M & 0 & 0.37 & 0.30  \\
\hline 2 & \multicolumn{2}{c|}{ 0.37$^*$ }& 0.03   \\
\hline 4 & 0.2 & 0 & $10^{-3}$  \\
\hline 6 & 0.22 & 0 & $10^{-5}$ \\
\hline
\end{tabular}
\caption{Computed fraction of peptide in the fibril phase compared
to the data of Kim {\em et al.} \cite{Kim:2004}.\\
$^*$ Aggregate fraction was observed to be a combination of fibrils
and oligomers.}
 \centering \label{tab:ureatable}
\end{table}

{\bf Electrostatics}

To compute $F_{es}$ we approximate the fibril as a smooth cylinder
of radius $R$ and uniform charge density.  The linear charge density
may be computed by noting that the average charge per $\beta$-strand
is $q/n_s$ and there are $2p$ strands per layer in the fibril.

To determine the peptide charge as a function of pH, we use
\begin{equation}
q=\sum_i^{\rm acidic\ residues} -\frac{10^{pH - pKa_i}}{1 + 10^{pH -
pKa_i}}+\sum_i^{\rm basic\ residues} +\frac{10^{pKa_i - pH}}{1 +
10^{pKa_i - pH}}, \label{eq:pHcharge}
\end{equation}
where the pKas of the amino acids are taken from Ref.
\cite{Sillero:1989}.  So, the charge density on each
peptide-molecule cylinder is $\rho=2qp/n_s a$, where $a=4.7$ \AA\ is
the spacing between amino acids.

\begin{figure}[ht]
\vspace{0.6 cm}
\begin{center}
\includegraphics[angle=0,scale=0.25]{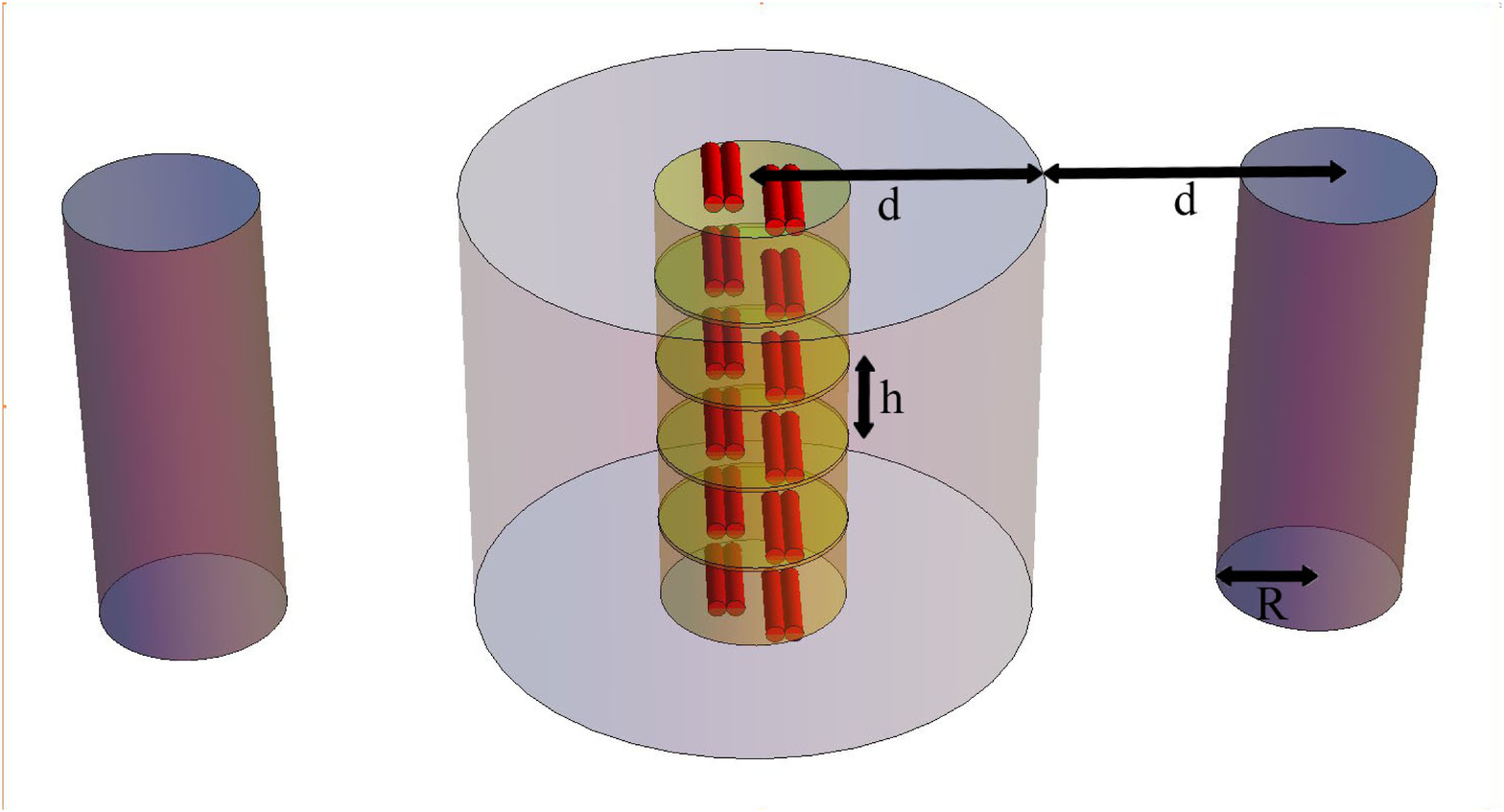}
\end{center}
\caption{ \label{fig:ESfig} Cylindrical fibril geometry used to
solve Eq. \ref{eq:PB}.  Fibrils are taken as cylinders of radius $R$
separated by a distance $2d$.  Red lines denote individual
$\beta$-strands (shown here for a fibril with $p=2$).  Horizontal
disks show the spacing $h$ between $\beta$-strand layers within the
fibril. Eq. \ref{eq:PB} is solved from surface of the reference
fibril (center), out to a distance $d$ (outer cylinder) with the
boundary condition $\psi'(d)=0$ reflecting the symmetry of the
electric potential between the cylinders.}
\end{figure}

We then solve for the electrostatic potential $\psi$ using the
Poisson-Boltzmann equation
\begin{equation}
\epsilon \frac{1}{r}\frac{\partial}{\partial r}r\frac{ \partial
\psi}{\partial r}=-e(c_+-c_-), \label{eq:PB}
\end{equation}
where the ion concentrations are $c_\pm=c_s e^{\mp e\psi/kT}$.  In
the linearized (Debye-Huckel) approximation, Eq. \ref{eq:PB} has the
solution
\begin{equation}
\psi(r) =  \frac{2 q}{2 \pi R a \epsilon \kappa {\rm K}_1 (\kappa
R)} {\rm K}_0(\kappa r),
\end{equation}
however, we use the numerical solution of the nonlinear Eq.
\ref{eq:PB} as the dimensionless potential $e\psi/kT$ can reach
values in excess of unity at the low salt concentrations we
consider.  Here $\kappa^{-1}$ is the Debye length, defined via
$\kappa^2 \equiv 2 e^2 c_0/ \left(\epsilon k_{\rm B}T\right)$.  At
infinite dilution the appropriate boundary conditions are
$\psi'(R)=-\rho/2\pi\epsilon R$ and $\psi(\infty)=0$, but for the
purposes of the numerical solver we employ the outer boundary
condition $\psi'(d)=0$ corresponding to a solution of fibrils
separated by an average distance $2d$.  We take $d=R+5\kappa^{-1}$
(see Fig. \ref{fig:ESfig}). For $d\gg\kappa^{-1}$ the influence of
the outer boundary condition will be minimal.

Once we have computed $\psi$, we get the electrostatic free energy
density of the peptide cylinder as \cite{Andelman:2005}
\begin{equation}
f=\frac{\epsilon}{2}\left(\frac{d \psi}{dr}\right)^2
+kT(c_+\ln(c_+/c_s) +c_-\ln(c_-/c_s)-c_+ -c_- + 2 c_s).
\label{eq:Fesdensity}
\end{equation}
The first term in Eq. \ref{eq:Fesdensity} is the electrostatic
energy stored in the electric field, and the remaining terms account
for the translational entropy of the ions in the screening layer.
We then compute the free energy per peptide using
\begin{equation}
\Delta F_{es}=2\pi\frac{n_s a}{2p}\int_R^\infty f(r)r dr
\label{eq:Fes}
\end{equation}
where $r$ is the radial coordinate perpendicular to the axis of the cylinder.

\end{document}